\documentstyle[emulateapj,epsf]{article}

\newcommand{\ts}{\thinspace}
\begin{document}

\submitted{T{\tiny HE} A{\tiny STRONOMICAL} J{\tiny OURNAL}, 120, No. 5 in press.}
\title{DEEP OPTICAL IMAGING OF A COMPACT GROUP OF GALAXIES, 
       SEYFERT'S SEXTET}
\author{\sc Shingo Nishiura$^{1,2}$, Takashi Murayama$^{1,2,3}$, 
        Masashi Shimada$^{1,2,4}$, Yasunori Sato$^{3,5}$,
        Tohru Nagao$^{1,3}$, Kohji Molikawa$^1$,
        Yoshiaki Taniguchi$^{1,2,3}$, and D. B. Sanders$^6$}

\affil{$^1$Astronomical Institute, Graduate School of Science, 
       Tohoku University, Aramaki, Aoba, Sendai 980-8578, Japan}
\affil{$^2$Visiting astronomer of Okayama Astrophysical Observatory, National
       Astronomical Observatory of Japan}
\affil{$^3$Visiting astronomer of Mauna Kea Observatories, University of Hawaii}
\affil{$^4$Asahi Optical Co., Ltd. Optcal Research Department, 
       2-36-9, Maeno-cho, Itabashi-ku, Tokyo, 174-8639, Japan}
\affil{$^5$ISAS, 3-1-1 Yoshinodai, Sagamihara, Kanagawa 229-8510, Japan}
\affil{$^6$Institute for Astronomy, University of Hawaii,
       2680 Woodlawn Drive, Honolulu, HI 96822}

\begin{abstract}
In order to investigate the dynamical status of Seyfert's Sextet (SS), we have
obtained a deep optical ($VR+I$) image of this group. Our image shows that
a faint envelope, down to a surface brightness $\mu_{\rm optical}$(AB)
$\simeq 27$ mag arcsec$^{-2}$, surrounds the member galaxies. 
This envelope is irregular in shape. It is likely that this shape
is attributed either to recent-past or to on-going galaxy interactions
in SS. If the member galaxies have experienced a number of
mutual interactions over a long timescale, the shape of the envelope
should be rounder. Therefore, the irregular-shaped morphology suggests
that SS is in an early phase of dynamical interaction among the member
galaxies. It is interesting to note that the soft X-ray image
obtained with ROSAT (Pildis et al. 1995) is significantly  similar 
in morphology.
We discuss the possible future evolution of SS briefly.
\end{abstract}



\section{INTRODUCTION}

Since groups of galaxies are intermediate in number density between
isolated galaxies and rich clusters of galaxies, it is important to
investigate their dynamical properties in detail.
Recently, the X-ray satellite ROSAT has been used to investigate 
the dark matter content of a large number of groups of galaxies because 
the hot gas probed in soft X-rays is generally believed to be 
gravitationally bound to the groups (Ponman \& Bertram 1993; Ebeling, 
Voges, \& B\"{o}hringer 1994; Pildis, Bregman, \& Evrard 1995; 
Saracco \& Ciliegi 1995; Ponman et al. 1996; Mulchaey \& Zabludoff 1998). 
As expected, the majority of groups of galaxies 
detected by ROSAT show round-shaped morphologies in the soft X-ray. 
On the other hand, a few groups such as Seyfert's Sextet (hereafter SS), 
show irregular-shaped soft X-ray morphologies (Pildis et al. 1995). 
The X-ray morphology can provide important information about the
dynamical status of groups of galaxies. 
Another useful method to
investigate the dynamical status is to search for a faint optical envelope
surrounding a group of galaxies (Nishiura 1998).
In this paper, in order to investigate the dynamical status of SS,
we present a very deep optical image of this group and compare it
with the soft X-ray halo obtained by Pildis et al. (1995).

SS is one of the  most famous as well as densest compact group of 
galaxies (Seyfert 1948a, 1948b). This group is also a Hickson Compact Group
of Galaxies, HCG 79 (Hickson 1982, 1993).  
The basic data of the galaxies in SS are listed in Table 1 taken from 
Hickson (1993). 
SS consists of four
redshift-accordant galaxies (HCG79a, HCG79b, HCG79c, and HCG79d). 
Note that HCG 79e is a redshift-discordant galaxy 
which is believed to have no physical relation to SS. Also, the NE optical 
fuzz is now considered to be tidal debris associated with HCG 79b (Rubin, 
Hunter, \& Ford 1991; Williams, McMahon, \& van Gorkom 1991; Mendes de 
Oliveira \& Hickson 1994; V\'{\i}lchez \& Iglesias-P\'{a}ramo 1998). 
We assume a distance to SS of 44$h^{-1}$ Mpc determined using the 
mean recession velocity of HCG 79a, HCG 79b, HCG 79c, 
and HCG 79d referenced to the galactic standard of rest, $V_{\rm GSR}$ = 4449 
km s$^{-1}$ (de Vaucouleurs et al. 1991), and a Hubble constant, 
$H_{0}$ = 100 $h$ km s$^{-1}$ Mpc$^{-1}$.


\begin{deluxetable}{cccccc}
\tablecaption{Properties of the member galaxies of SS and the discordant galaxy
HCG 79e}
\tablehead{
\colhead{Name} & \colhead{Type}    & \colhead{$V_{\rm GSR}$}
               & \colhead{$B_{\rm T}^{0}$} & \colhead{$L_{\rm B}$}
               & \colhead{$L_{\rm X, exp}$} \nl
\colhead{} & \colhead{}    & \colhead{(km s$^{-1}$)}
               & \colhead{(mag.)}  & \colhead{($h^{-2}$ ergs s$^{-1}$)}
               & \colhead{($h^{-2}$ ergs s$^{-1}$)} \nl
}
\startdata
HCG 79a & E0  & 4395  & 14.35 & 43.32 & 38.86 \nl
HCG 79b & S0  & 4547  & 13.78 & 43.55 & 39.43 \nl
HCG 79c & S0  & 4247  & 14.72 & 43.18 & 38.50 \nl
HCG 79d & Sdm & 4604  & 15.87 & 42.72 & 38.50 \nl
HCG 79e & Scd & 19910 & 15.87 & 44.03 & 40.09 \nl
\enddata
\end{deluxetable}

\section{OBSERVATIONS}

\subsection{Optical Imaging}

Optical $VR$- and $I$-band deep images of SS were obtained with 
the 8192 $\times$ 8192 (8K) CCD mosaic camera (Luppino et al. 1996) 
attached to the f/10 Cassegrain focus of the University of Hawaii 
2.2 m telescope at Mauna Kea Observatory, on 20 May 1999 ($VR$) 
and 23 May 1999 ($I$).
The camera provided a $\sim 18\arcmin \times 18\arcmin$ field 
of view. Two-pixel binning was used yielding a spatial resolution 
of 0.26 arcsec per element. 
The integration time for each exposure was set to 8 minutes. 
Twenty-three exposures for the $VR$-band and 24 exposures for 
the $I$-band were taken; thus, the total integration time amounted 
to 10,560 seconds in the $VR$-band image and 11,520 seconds 
in the $I$-band image.
The seeing was $\sim$0.8 arcsec during the observation.

Data reduction was performed in a standard way using IRAF\footnote{
Image Reduction and Analysis Facility (IRAF) is distributed 
by the National Optical Astronomy Observatories, which are operated 
by the Association of Universities for Research in Astronomy, 
Inc., under cooperative agreement with the National Science Foundation.}. 
Flux calibration was made using the data of photometric standard 
stars in the field of SA{\ts}103, SA{\ts}104, SA{\ts}107, SA{\ts}109, 
and SA{\ts}110 (Landolt 1992). 
Since the $VR$-band is not a standard photometric band (Jewitt, Luu \& 
Chen 1996), we adopted an AB magnitude scale for this bandpass. 
The photometric errors were estimated to be $\pm 0.05$ mag for 
the $VR$-band and $\pm 0.03$ mag for the $I$-band. 
The limiting surface brightnesses are $\mu^{\rm lim}_{VR}$ = 28.7 mag 
arcsec$^{-2}$ and $\mu^{\rm lim}_{I}$ = 28.1 mag arcsec$^{-2}$, 
corresponding to a 1 $\sigma$ variation in the background. 
The $VR$, $I$, and $VR+I$ images are shown in Figure 1 (a), 
Figure 2 (a), and Figure 3, respectively. 

\begin{center}
\epsscale{0.5}
\figcaption{
(a) $VR$ image of SS. Black contours are drawn above $\mu_{AB}=
27$ mag arcsec$^{-2}$ with an interval of 1 mag arcsec$^{-2}$.
(b) Images of model galaxies of SS in $VR$-band. 
And, (c) Model subtracted image of SS in $VR$-band. 
The holizontal bars correspond to 1 arcmin $\simeq$ 12.8 $h^{-1}$ kpc.
North is up and east is to the left.
\label{fig1}
}
\end{center}

\begin{center}
\epsscale{0.5}
\figcaption{
(a) $I$ image of SS. Black contours are drawn above $\mu_{I}=
27$ mag arcsec$^{-2}$ with an interval of 1 mag arcsec$^{-2}$.
(b) Images of model galaxies of SS in $I$-band. 
And, (c) Model subtracted image of SS in $I$-band. 
The holizontal bars correspond to 1 arcmin $\simeq$ 12.8 $h^{-1}$ kpc.
North is up and east is to the left.
\label{fig2}
}
\end{center}

\subsection{Optical Spectroscopy}

In order to investigate the nuclear activity of the member galaxies of SS,
we have performed optical spectroscopy during the course of our optical 
spectroscopy program of Hickson compact groups of galaxies 
(Shimada et al. 2000; Nishiura et al. 2000). 
Optical spectra were obtained with the Cassegrain spectrograph attached
to the 188 cm telescope at the Okayama Astrophysical Observatory (OAO)
on 22 February (HCG 79a), 15 August (HCG 79b), and 18 August 1996
(HCG 79c and HCG 79d).
A 600 lines/mm grating blazed at 7,500 \AA\ was used. 
The spectral coverage was 6,200 \AA~ -- 6,900 \AA~ with a 
spectral resolution of 2.4 \AA~ at 6,500 \AA.
The integration time was 1800 seconds for each galaxy.
The spatial resolution was 1.75 arcsec pixel$^{-1}$ and the
slit width was 1.8 arcsec. The seeing size was 2 arcsec during the
observations. The data reduction was made using IRAF with a special 
package SNGRED for the OAO spectrograph (Kosugi et al. 1995).
Spectroscopic standard stars (HD 84937, HD 161817, HD 217086,
BD 322642, and  Feige 15) were observed to calibrate the spectra.
One-dimensional nuclear spectra of the four member galaxies
were extracted by tracing the central
two pixels; i.e., 3.5 arcsec.
The final nuclear spectra are shown in Figure 4.

\begin{center}
\epsscale{0.75}
\figcaption{
(a) $I$ image of SS. Black contours are drawn above $\mu_{I}=
27$ mag arcsec$^{-2}$ with an interval of 1 mag arcsec$^{-2}$.
(b) Images of model galaxies of SS in $I$-band. 
And, (c) Model subtracted image of SS in $I$-band. 
The holizontal bars correspond to 1 arcmin $\simeq$ 12.8 $h^{-1}$ kpc.
North is up and east is to the left.
\label{fig3}
}
\end{center}

\begin{center}
\epsscale{0.85}
\plotone{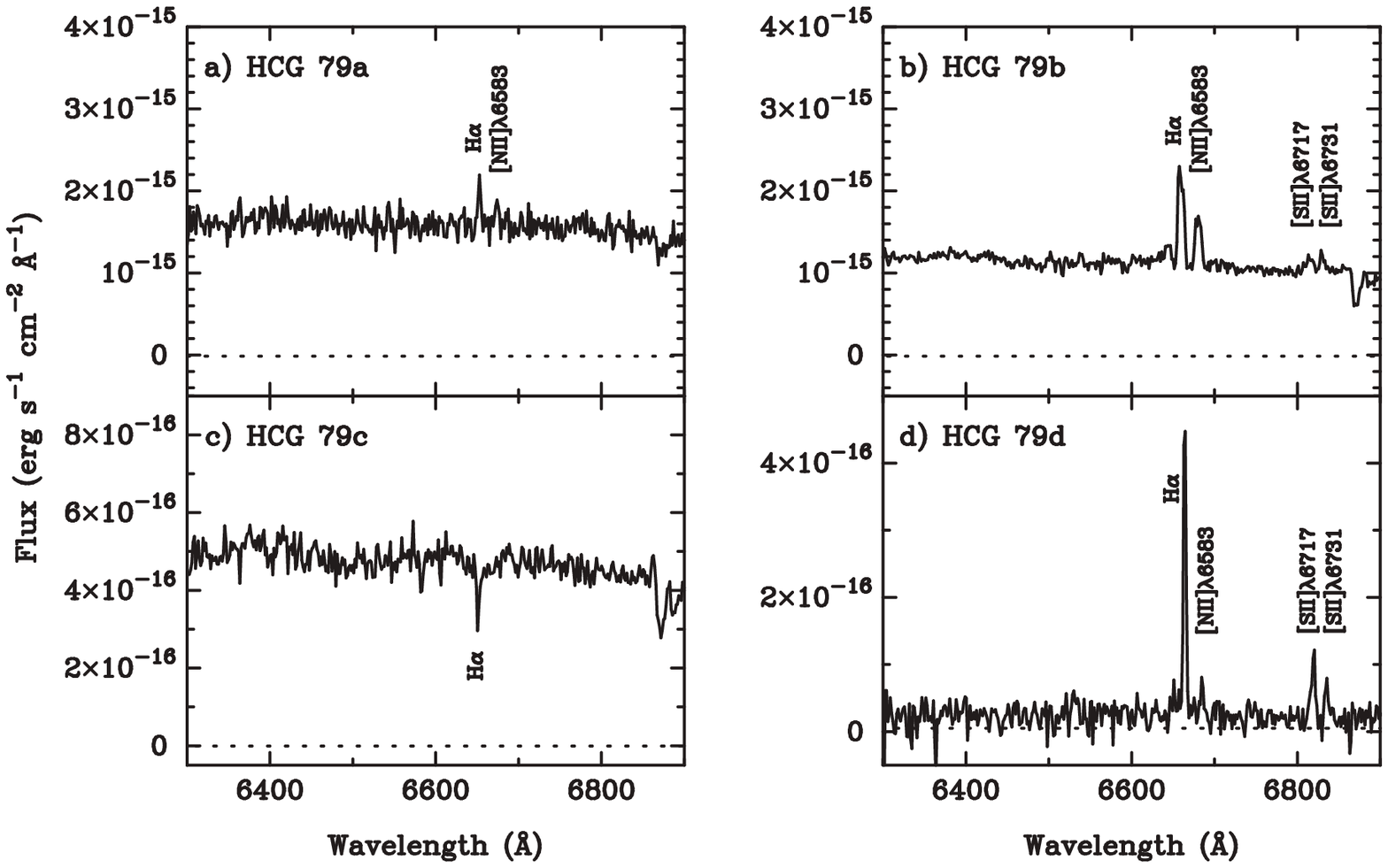}
\figcaption{Optical spectra of a$)$ HCG 79a, b$)$ HCG 79b, 
 c$)$ HCG 79c, and d$)$HCG 79d.
\label{fig4}
}
\end{center}


\begin{deluxetable}{ccccccc}
\tablecaption{Parameter of model galaxies}
\tablehead{
\colhead{}     & \colhead{}            & \colhead{radius} 
               & \colhead{}            & \colhead{} 
               & \colhead{}            & \colhead{P.A.}   \nl
\colhead{Name} & \colhead{band}        & \colhead{(arcsec)}
               & \colhead{mag.}        & \colhead{profile\tablenotemark{a}}
               & \colhead{ellipticity} & \colhead{(degree)} 
}
\startdata
HCG 79a      & $VR$ & 7.98 & 13.09 & $r^{1/4}$ & 0.30 & 158.0 \nl
             & $I$  & 7.72 & 12.30 & $r^{1/4}$ & 0.30 & 158.0 \nl
HCG 79b      & $VR$ & 3.42 & 12.92 & exp       & 0.67 & 177.0 \nl
             & $I$  & 3.24 & 12.65 & exp       & 0.65 & 174.0 \nl 
HCG 79c      & $VR$ & 2.59 & 13.65 & $r^{1/4}$ & 0.52 & 126.0 \nl
             & $I$  & 3.46 & 12.99 & $r^{1/4}$ & 0.52 & 126.0 \nl 
HCG 79d      & $VR$ & 8.09 & 15.80 & exp       & 0.74 &  87.0 \nl
             & $I$  & 8.48 & 15.11 & exp       & 0.77 &  86.0 \nl
HCG 79e      & $VR$ & 3.90 & 15.24 & exp       & 0.06 &  54.0 \nl
             & $I$  & 3.84 & 14.55 & exp       & 0.06 &  58.0 \nl
Tidal debris & $VR$ & 6.35 & 17.24 & exp       & 0.88 & 133.4 \nl
             & $I$  & 6.09 & 16.49 & exp       & 0.90 & 133.4 \nl
\enddata
\tablenotetext{a}{$r^{1/4}$ means de Vaucouleurs' $r^{1/4}$ law 
                  and exp means exponential law.}
\end{deluxetable}

\section{RESULTS AND DISCUSSION}

\subsection{The Morphology of the Faint Optical Envelope}

As shown in Figure 1 (a), Figure 2 (a), and Figure 3, 
we can see a faint optical envelope surrounding SS both in the $VR$ 
image and in the $I$ image. 
However, since the member galaxies are very close together 
on the sky, we cannot rule out a possibility 
that we may misidentify the overlapping isophotes of galaxies for 
the real faint optical envelope. 
In order to confirm the presence of the faint optical envelope, 
it is necessary to subtract off models of all the galaxies in SS. 
For this purpose we have made a model of each galaxy in SS 
by fitting a surface brightness profile of its outer region. 
In this procedure, we adopt a de Vaucouleurs' 
$r^{1/4}$ law for HCG 79a and 79c and an exponential law 
for HCG 79b, 79d, 79e. We also made a model for the tidal debris 
east of HCG 79b (see Fig. 3) with an exponential law. 
Surface photometry was carried out using the Surface Photometry 
Interactive Reduction and Analysis Library (Hamabe \& Ichikawa 1992). 
We made an image of the model SS by adding the above models. 
In this procedure we use the ARTDATA package with the fitting parameters 
derived by the surface photometry. In Table 2 we list the parameters 
of our model galaxies. 
For the model of the tidal debris east of HCG 79b, 
it seems that there is a much better fit would be found 
with a rounder model at a larger position angle. 
Fainter parts of the debris may be connected to the optical envelope. 
Therefore, we have made our model fit for the brighter part of 
the debris and then subtracted it from the image. 

\begin{center}
\epsscale{0.85}
\plotone{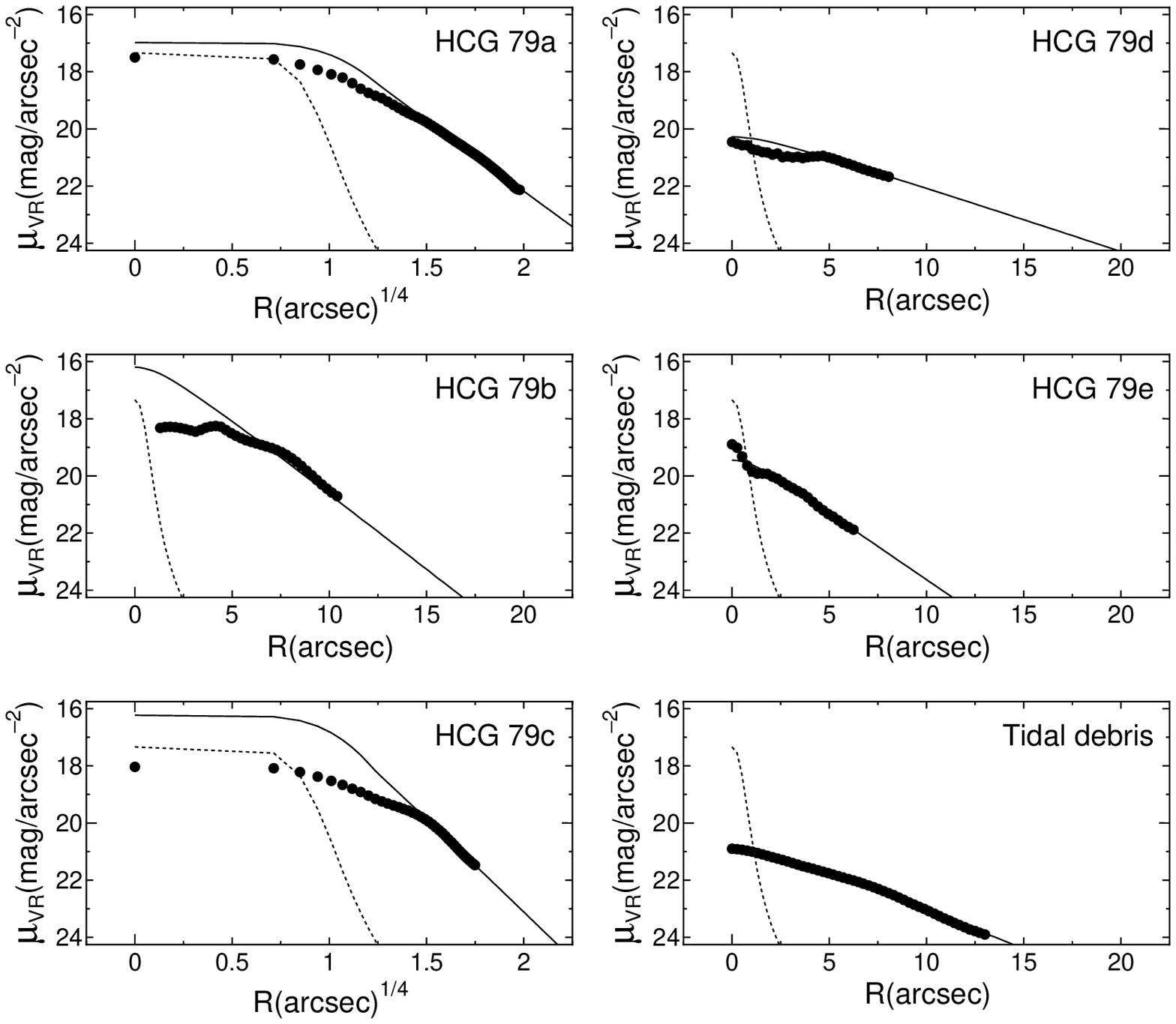}
\figcaption{
$VR$-band surface brightness profiles along the major axis of the 
galaxies in SS. The filled circles indicate the profiles of observed 
galaxies. The solid lines indicate the profiles of model galaxies. 
The dotted lines indicate the profiles of a star observed in the same 
frame. 
\label{fig5}
}
\end{center}

\begin{center}
\epsscale{0.85}
\plotone{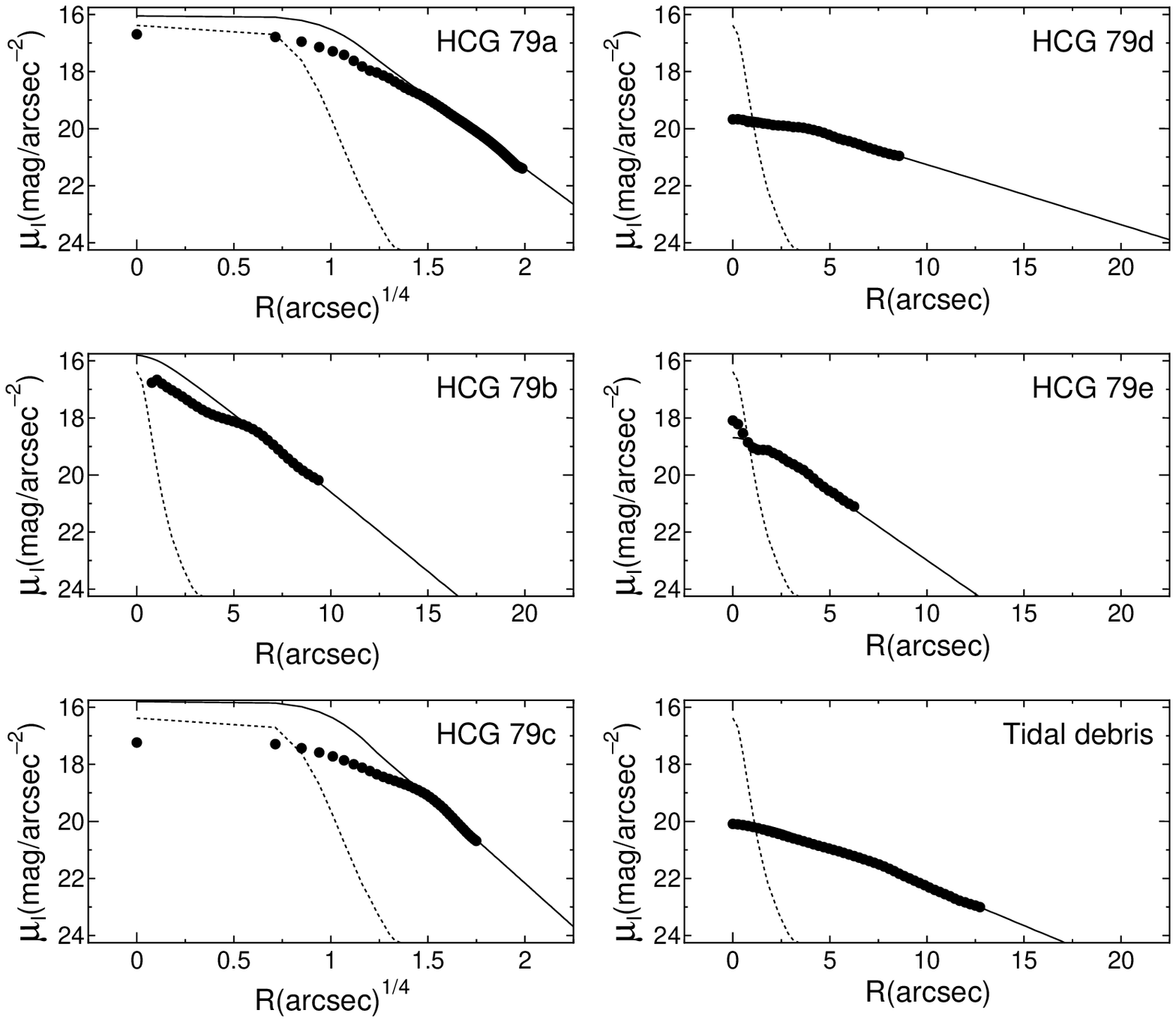}
\figcaption{
$I$-band surface brightness profiles along the major axis of 
galaxies in SS. The filled circles indicate the profiles of observed 
galaxies. The solid lines indicate the profiles of model galaxies. 
The dotted lines indicate the profiles of a star observed in the same 
frame.
\label{fig6}
}
\end{center}


\begin{deluxetable}{cccccc}
\tablecaption{Emission-line properties of the member galaxies of SS}
\tablehead{
\colhead{Name} & \colhead{log $f_{\rm H\alpha}$}
               & \colhead{[N {\sc ii}]$\lambda$6583/H$\alpha$} 
               & \colhead{[S {\sc ii}]$\lambda$6583/H$\alpha$}
               & \colhead{FWHM(H$\alpha$)} 
               & \colhead{EW(H$\alpha$)}\nl
\colhead{}     & \colhead{(ergs s$^{-1}$ cm$^{-2}$)}  
               & \colhead{}
               & \colhead{}  
               & \colhead{(km s$^{-1}$)}
               & \colhead{(\AA)}\nl
}
\startdata
HCG 79a & $-14.60\pm0.09$ & $0.52\pm0.19$ & \nodata 
        & $150\pm37$      & $-1.13^{+0.21}_{-0.26}$ \nl
HCG 79b & $-14.00\pm0.02$ & $0.53\pm0.04$ & $0.27\pm0.03$ 
        & $337\pm14$      & $-6.76^{+0.30}_{-0.31}$ \nl
HCG 79c & absorption      & \nodata       & \nodata 
        & \nodata         & \nodata    \nl
HCG 79d & $-14.75\pm0.02$ & $0.12\pm0.02$ & $0.44\pm0.06$ 
        & $121\pm7$       & $-116\pm5$ \nl
\enddata
\end{deluxetable}
 
The results are shown in Figure 1 (b) ({\it VR}-band) and 
2 (b) ({\it I}-band). The model subtracted images are shown in 
Figure 1 (c) ({\it VR}-band) and 2 (c) ({\it I}-band). 
Figure 5 and 6 show the {\it VR}- and {\it I}-band surface brightness 
profiles of all the galaxies in SS. The filled circles indicate 
the observed profiles and solid lines indicate the profiles of 
the model galaxies.      
As a result of the model subtraction, we can still see 
the faint optical envelope 
surrounding SS both in the $VR$ image and in the $I$ image.

Next, in order to check whether the skirts of the point 
spread function (PSF) of the galactic nuclei contaminate 
the faint optical envelope, 
we compared the surface brightness profiles of all the galaxies in SS 
with that of a star in the same frame of SS. The surface 
brightness profile of the star is shown as dotted lines in Figure 5 and 6. 
We cannot find any contribution of the light of the PSF to the faint 
optical envelope surrounding SS both in the $VR$ image and in the $I$ image. 
Throughout these two checks we have confirmed that the faint optical 
envelope is really present around SS. 

Using the images in Figure 1c and Figure 2c, we roughly calculate 
the {\it VR} and {\it I} magnitudes of the faint optical envelope 
by integrating the light without any point sources 
(i.e., stars in our Galaxy) 
within a circle of the centered at  
$\alpha = 15^{\rm h} 56^{\rm m} 59\fs6$, and 
$\delta = +20^{\rm d} 54^{\rm m} 59\fs6$ 
with a radius of 70 arcsec enclosing the optical faint envelope of SS. 
The regions which appear black on Figure 1c and 2c are also ignored. 
We obtain $VR_{\rm AB} \simeq 14.5$ and $I \simeq 13.7$ 
for the envelope. The luminosity contribution of the envelope to 
the total luminosity is $\approx$ 12 \% in {\it VR} and $\approx$ 
13 \% in {\it I}. This implies that the envelope consists of 
red stars. 
If we assume that the mass-to-light ratio of the stellar component of 
the optical faint envelope is $M/L=3.57\pm 0.43$ 
($M_{\odot}/L_{\odot}$) in {\it I}-band (Worthey 1994), 
we obtain the mass of the optical faint envelope,
$M_{\rm env} \simeq (9.1\pm 1.1)\times 10^{9} M_{\odot}$.  

The morphology of the faint optical
envelope in the $VR$ image is quite similar to that in the $I$ image.
The overall morphology of the individual galaxies is consistent with 
that obtained previously (Sulentic \& Lorre 1983).
The faint optical envelope shows an irregular shape. It is likely that this 
shape is attributed either to recent-past or to on-going galaxy interactions
in SS. If the member galaxies have experienced a number of
mutual interaction over a long timescale, the shape of the envelope
might reasonably be expected to be rounder. Therefore, the irregular-shaped 
morphology suggests that SS is in an early phase of dynamical interactions
among the member galaxies.

Rubin et al. (1991) and Nishiura et al. (2000) showed that the optical 
rotation curve of HCG 79d is peculiar compared to those of normal spirals. 
Williams et al. (1991) obtained an H {\sc i} map of SS and found that
the H {\sc i} emission, whose intensity peak appears at the center of 
HCG 79d, shows a southern extension from HCG 79d. They interpreted this
as possible evidence for the interaction between HCG 79d and 79b.
It is thus strongly suggested that HCG 79d shows 
tidal effects from this interaction. 

\begin{center}
\epsscale{0.75}
\figcaption{
The soft X-ray image (black contours) taken from Pildis et al. 
(1995) is overlaid on our $(VR+I)$ CCD image. The Soft X-ray contours 
are drawn at 1$\sigma$, 2$\sigma$, 4$\sigma$, and 8$\sigma$ above the 
background level.
\label{fig7}
}
\end{center}

Finally, we compare our very deep optical image [$VR+I$;
Figure 3] with the soft X-ray image taken by Pildis et al.
(1995) in Figure 7.
It is shown that the soft X-ray morphology is quite similar
to that seen in the optical.   
The optical faint envelope is most likely attributed to stars liberated from
the member galaxies through past and on-going tidal interactions.
Therefore, the morphological similarity between the optical and soft X-ray
images implies that the dark matter in SS was originally 
associated 
with the individual galaxies and then has been liberated tidally
together with the stars.

\subsection{The Origin of Soft-X Ray Emission}

The total soft X-ray luminosity of SS is given only as an upper limit,  
$L_{\rm X, tot} < 1.3 \times 10^{41} h^{-2}$ ergs s$^{-1}$ 
with ROSAT PSPC (Ponman et al. 1996; hereafter P96). On the other hand, using the same 
data, Pildis et al. (1995) detected soft X-ray photons at the $2.6 \sigma$ 
level. The net count of the extended soft X-ray emission was 
28.4$\pm$11.1 during the total integration time of 4910{\ts}sec.
Using the software Xspec\footnote{Xspec is available 
at http://heasarc.gsfc.nasa.gov/webspec/webspec.html},
we estimate the soft X-ray luminosity of the extended emission;
$L_{\rm X} \sim 1.3 \times 10^{40} h^{-2}$ ergs s$^{-1}$
for $kT$ = 0.5 keV or $L_{\rm X} \sim 1.4 \times 10^{40} h^{-2}$ 
ergs s$^{-1}$ for $kT$ = 1 keV, where the metal abundances are 
assumed to be 0.3 times solar. 
It is difficult to resolve X-ray emission from individual galaxies 
using the ROSAT PSPC because of the compactness of SS itself, 
even if the exposure time were long enough.
It is possible that a part of the X-ray emission may be associated 
with individual member galaxies.  
Here we also attempt to examine whether or not the soft X-ray 
emission of SS really arises from hot gas since supernova remnants, 
low- and high-mass X-ray binaries, and/or active galactic nuclei could 
also contribute substantially to the soft X-ray emission to some extent 
if they are present in the member galaxies. 

First we estimate the expected soft X-ray luminosities of discrete
sources such as X-ray binaries in the individual galaxies $L_{\rm X, gal}$.
Such soft X-ray luminosities can be estimated using the following empirical 
$L_X$ - $L_B$ relationship for galaxies where $L_B$ is the absolute blue
luminosity. 
\[\log L_{\rm X} = (2.47 \pm 1.01) \times \log L_{\rm B} - (68.36 \pm 44.08),\]
for early-type galaxies (Brown \& Bregman 1998), and
\[\log L_{\rm X} = (1.21 \pm 0.03) \times \log L_{\rm B} - (13.29 \pm 1.39).\]
for late-type galaxies (Read, Ponman, \& Strickland 1997; Vogler, Pietish, \&
Kahabka 1996).
In Table 1, we give a summary of the expected soft X-ray luminosities for 
the individual galaxies where $L_B$ is estimated using the apparent blue
magnitudes (Hickson 1993).
Although HCG 79e is the redshift-discordant galaxy,
this galaxy also contributes to the observed soft X-ray luminosity of SS because
it lies in a line of sight toward the group. 
We obtain an expected total soft X-ray luminosity coming from individual soft 
X-ray sources in the five galaxies; 
$L_{\rm X, gal} \simeq 4.4 \times 10^{39}h^{-2}$ ergs s$^{-1}$, which is 
much lower than the observed soft X-ray luminosity of SS.

Second, we estimate possible contributions of soft X-ray emission
from starburst phenomena in the member galaxies, $L_{\rm X, H{\sc ii}}$.
The spectroscopic properties of the member galaxies based on 
our optical spectra (Figure 4) are summarized in Table 3. 
Our results show that HCG 79a, 79b and 79d have a weak H{\sc ii} 
nucleus. HCG 79c shows no emission-line activity.
For starburst galaxies, an average soft X-ray to 
optical $B$ luminosity ratio is estimated as, 
$\log L_{\rm X}/L_{B} \simeq -3.87 \pm 0.19$ (Read et al. 1997). 
We estimate that the soft X-ray luminosities from an \ion{H}{2} 
nucleus in HCG 79a, 79b and 79d 
are $L_{\rm X, H{\sc ii}} = 2.8 \times 10^{39} h^{-2}, 
4.8 \times 10^{39} h^{-2}$, and $7.1 \times 10^{38} h^{-2}$ ergs s$^{-1}$, 
respectively. Thus, we obtain a total soft X-ray luminosity from 
starbursts in these galaxies, $L_{\rm X, H{\sc ii}}$ = 
$8.3 \times 10^{39} h^{-2}$ ergs s$^{-1}$

The model estimated total soft X-ray luminosity from SS, which is 
the sum total we have calculated above for the soft X-ray luminosity of 
the individual galaxies, is 
$1.3 \times 10^{40}$ ergs s$^{-1}$. 
This value is nearly identical to our estimated value using Xspec,  
$L_{\rm X,tot}=1.3\times 10^{40} h^{-2}$ergs s$^{-1}$ 
with $kT$=0.5keV and $Z$=0.3 $Z_{\odot}$ assumed, and 
$L_{\rm X,tot}=1.4\times 10^{40} h^{-2}$ergs s$^{-1}$ 
with $kT$=1.0keV and $Z$=0.3 $Z_{\odot}$ assumed. 

Finally, we estimate the X-ray luminosity of hot gas 
trapped by the gravitational 
potential of SS, $L_{\rm X, gas}$.
Mulchaey \& Zabludoff (1998) find a correlation between 
soft X-ray luminosity from hot gas within poor clusters 
$L_{\rm X, gas}$ and radial velocity dispersion $\sigma_{\rm r}$ 
of galaxies in them, 
\[\log L_{\rm X} = (31.61 \pm 1.09) + \log h^{-2} 
                 + (4.29 \pm 0.37) \times \log \sigma_{\rm r}.\]
Using this empirical relation, 
we obtain $L_{\rm X, gas} = 6.2 \times 10^{40}h^{-2}$ ergs s$^{-1}$ 
using the observed value $\sigma_{\rm r} = 138$ km s$^{-1}$ for SS.
This value is a factor of five larger than the observed 
X-ray luminosity. 
Since the sum total we calculate for the soft X-ray luminosity 
of the individual galaxies is high enough to explain the observed 
X-ray luminosity, it is not necessarily to take account of the
mass contribution from the hot gas. 
Deeper X-ray observations of SS are needed to discuss the nature of 
the hot gas in SS. 

\subsection{The Dynamical Properties of Seyfert's Sextet}

To discuss the dynamical properties of SS, we first estimate the 
dynamical mass of the system. Heisler, Tremaine 
\& Bahcall (1985) have proposed four mass estimators for small-member 
galaxy associations such as compact groups of galaxies; 1) the virial mass 
estimator $M_{\rm vir}$, 2) the projected mass estimator $M_{\rm proj}$,
3) the average mass estimator $M_{\rm avg}$, and 4) the median mass 
estimator $M_{\rm med}$. In order to 
avoid  accidental errors in each mass estimator, we estimate the 
total mass of SS by averaging masses derived from these four estimators;
$M_{\rm tot} \simeq 3.6 \times 10^{11}h^{-1} M_{\odot}$.
The results are summarized in Table 4.

\begin{center}
\begin{center}
T{\scriptsize ABLE} 4\\
P{\scriptsize ROPERTIES OF} SS
\\
\end{center}
\begin{minipage}{8cm}
\footnotesize
\begin{tabular}{lr}
\hline\hline
Properties & \\
\hline
$L_{\rm X, tot}$ (P96)  & $< 1.3 \times 10^{41}$ $h^{-2}$ ergs s$^{-1}$ \\   
$L_{\rm X, tot}$ ($kT$=0.5 keV, $Z$=0.3$Z_{\odot}$)  
                        & $  1.3 \times 10^{40}$ $h^{-2}$ ergs s$^{-1}$ \\
$L_{\rm X, tot}$ ($kT$=1.0 keV, $Z$=0.3$Z_{\odot}$)  
                        & $  1.4 \times 10^{40}$ $h^{-2}$ ergs s$^{-1}$ \\ 
$L_{\rm X, gal}$       & $4.4  \times 10^{39}$ $h^{-2}$ ergs s$^{-1}$ \\
$L_{\rm X, H{\sc ii}}$   & $8.3  \times 10^{39}$ $h^{-2}$ ergs s$^{-1}$ \\
$L_{\rm X, gas}$        & $6.2  \times 10^{40}$ $h^{-2}$ ergs s$^{-1}$ \\
$L_{\rm B}$             & $2.0  \times 10^{10}$ $h^{-2}$ $L_{\odot}$   \\ 
$M_{\rm vir}$

\footnote{
Hickson et al. (1992) obtained a virial mass of SS, 
$M_{\rm vir} = 1.38 \times 10^{11} M_{\odot}$, 
using $M_{\rm vir} \simeq \pi V^{2}R/2G$, where $V$ is the estimated 
intrinsic three-demensional velocity dispersion, 
$V = [3(<v^{2}>-<v>^{2})-<\delta v^{2}>]^{1/2}$, $v$ is the measured 
radial velocity of the galaxy, $\delta v$ is the estimated velocity 
error, and $<>$ denotes the average over all galaxies in the group, 
$R$ is the median length of the two-dimensional galaxy-galaxy separation, 
$G$ is the gravitational constant. In this study 
we estimated the virial mass of SS using the virial mass estimator 
taken from Heisler et al. (1985), 
$M_{\rm vir} = [3 \pi N \sum_{i}(v_{i}-<v>)^{2} / 2G 
\sum_{i<j}(1/R_{ij})]$, where $N$ is the number of member 
galaxies, $v_{i}$ is the radial velocity of the $i$-th galaxy, 
$R_{ij}$ is the two-dimensional distance between the $i$-th galaxy 
and the $j$-th galaxy.
}           
                        & $(3.89^{+1.18}_{-3.10}) \times 10^{11}h^{-1}
M_{\odot}$   \\
$M_{\rm proj}$          & $(3.63^{+9.88}_{-2.23}) \times 10^{11}h^{-1}
M_{\odot}$   \\
$M_{\rm avg}$           & $(3.80^{+9.79}_{-2.13}) \times 10^{11}h^{-1}
M_{\odot}$   \\
$M_{\rm med}$           & $(3.09^{+3.76}_{-2.02}) \times 10^{11}h^{-1}
M_{\odot}$   \\
$M_{\rm tot}$           & $(3.63^{+5.41}_{-2.43}) \times 10^{11}h^{-1}
M_{\odot}$   \\
$M_{\rm tot}/L_{\rm B}$ & $(18^{+27}_{-12})h M_{\odot}/L_{\odot}$        
 \\ 
$M_{\rm gal}$           & $(2.67\pm 0.11) \times 10^{11}$ $h^{-1}M_{\odot}$ \\
\hline

\end{tabular}
\normalsize
\end{minipage}
\end{center}

Next, we estimate the mass including dark matter associated with
the member galaxies of SS, $M_{\rm gal}$. The mean mass-to-luminosity ratios 
for each Hubble type are as follows: 13.7 $\pm$ 0.3 for early-type galaxies
(Bacon, Monnet, \& Simien 1985), and 9.1 $\pm$ 4.0 for Sdm galaxies, where 
we have estimated a typical mass-to-luminosity ratio for Sdm galaxies by averaging
that of Sc galaxies, 5.2 $\pm$ 0.4 (Rubin et al. 1985), and
that of irregular galaxies, 12.9 $\pm$ 7.5 (Lo, Sargent \& Young 1993).
Although HCG 79a and 79b are LINERs, the line emission has little
effect on their $B$-band luminosities because the equivalent widths of
H$\alpha$ emission, $EW$(H$\alpha$), of these galaxies are less than  
several \AA ~ (see Table 3).
HCG 79d has a much larger $EW$(H$\alpha$).
However, its contribution to the total $B$-band luminosity of SS
is only a few percent. Therefore, the $B$ luminosity
derived above can be regarded as the stellar $B$ luminosity.
We thus obtain an estimate of the total (i.e., stars, gas, and the dark matter)
mass of the individual galaxies of SS,
$M_{\rm gal} \simeq 2.7 \times 10^{11} h^{-1} M_{\odot}$;
note that $M_{\rm gal}$ does not contain the mass of dark matter
in intragroup space. 
Since this mass is {\ts}73$\pm 3${\ts}\% of the estimated dynamical 
mass of SS,
it would appear that the majority of the dark matter may still be associated 
with the member galaxies. 
Since the dynamical mass measures the mass inside the circle enclosing 
the galaxies and is therefore insensitive to any mass outside it, 
we note that the ratio of 73\% is an upper limit. 

\begin{table*}
\begin{center}
T{\scriptsize ABLE} 5\\
B{\scriptsize ASIC PROPERTIES OF THE ELLIPTICALS WITH LARGE-SCALE} 
X-{\scriptsize RAY HALO AND} SS
\\
\begin{tabular}{cccccc}

\hline\hline
               & $R$          
               & $M_{\rm tot}$
               & $B_{T}^{0}$
               & $L_{\rm B}$  
               & $M_{\rm tot}/L_{\rm B}$ \\
          Name & (kpc)        
               & ($M_{\odot}$)
               & (mag.)
               & ($L_{\odot}$) 
               & ($M_{\odot}/L_{\odot}$) \\
\hline
SS                &   8$h^{-1}$
                  & 3.63$\times 10^{11} h^{-1}$
                  & 12.94
                  & 2.0$\times 10^{10} h^{-2}$
                  & 18$h$ \\

RXJ 1340.6$+$4018 & 170$h^{-1}$ 
                  & 1.4$\times 10^{13} h^{-1}$ 
                  & \nodata 
                  & $3.7\times 10^{10} h^{-2}$\tablenotemark{a} 
                  & 373$h$ \\
NGC 1132          & 243$h^{-1}$ 
                  & 1.9$\times 10^{13} h^{-1}$ 
                  & 13.03\tablenotemark{b}
                  & $4.6\times 10^{10} h^{-2}$ 
                  & 413$h$ \\
NGC 4636          & 300$h^{-1}$ 
                  & 9$\times 10^{12} h^{-1}$   
                  & 10.43\tablenotemark{b}
                  & $3.0\times 10^{10} h^{-2}$ 
                  & 297$h$ \\ 
\hline
\end{tabular}
\end{center}
\tablenotetext{a}{Estimated from $M_{V_{T}}=-23.5$ (Ponman et al. 1994) 
                  with adopting $B-V=1.04$ (Arimoto \& Yoshii 1987) 
                  and $H_{0}$=100$h$ km s$^{-1}$ Mpc$^{-1}$.}
\tablenotetext{b}{Taken from de Vaucouleurs et al. (1991).}
\end{table*}

\subsection{Dynamical Evolution of Seyfert's Sextet}

We discuss the possible future evolution of SS.
The irregular-shaped soft X-ray morphology of SS suggests that the dark
matter in SS has not yet completely relaxed dynamically. 
This appears consistent with the following observations:
1) the member galaxies in SS show morphological and kinematic
evidence for violent interactions, and 2) the majority of the dynamical 
mass of SS can be associated with the member galaxies.
It is thus strongly suggested that SS is  
a dynamically-young, compact group of galaxies which will eventually merge.
It has been numerically shown that the merging time scale of a
compact group of galaxies, each member of which has a dark halo, is much 
shorter (i.e., $\sim 10^{9}$ yr) than the Hubble time (Athanassoula, 
Makino, \& Bosma 1998).
Therefore, it is expected that SS will most likely merge into a single object 
within a time scale less than $10^9$ years and finally evolve into an
elliptical galaxy with a large, virialized halo.
Such a fossil group of galaxies has been reported;
a soft X-ray source RXJ 1340.6+4018 turns out to be an elliptical galaxy
with a large-scale ($\sim$ 250 $h^{-1}$ kpc)  X-ray halo (Ponman et
al. 1994; Jones et al. 2000).
Other candidates may be the elliptical galaxy
NGC 4636 in the Virgo cluster of galaxies (Matsushita et al. 1998) and
an isolated elliptical galaxy NGC 1132 (Mulchaey \& Zabludoff 1999).

\begin{center}
\epsscale{0.75}
\plotone{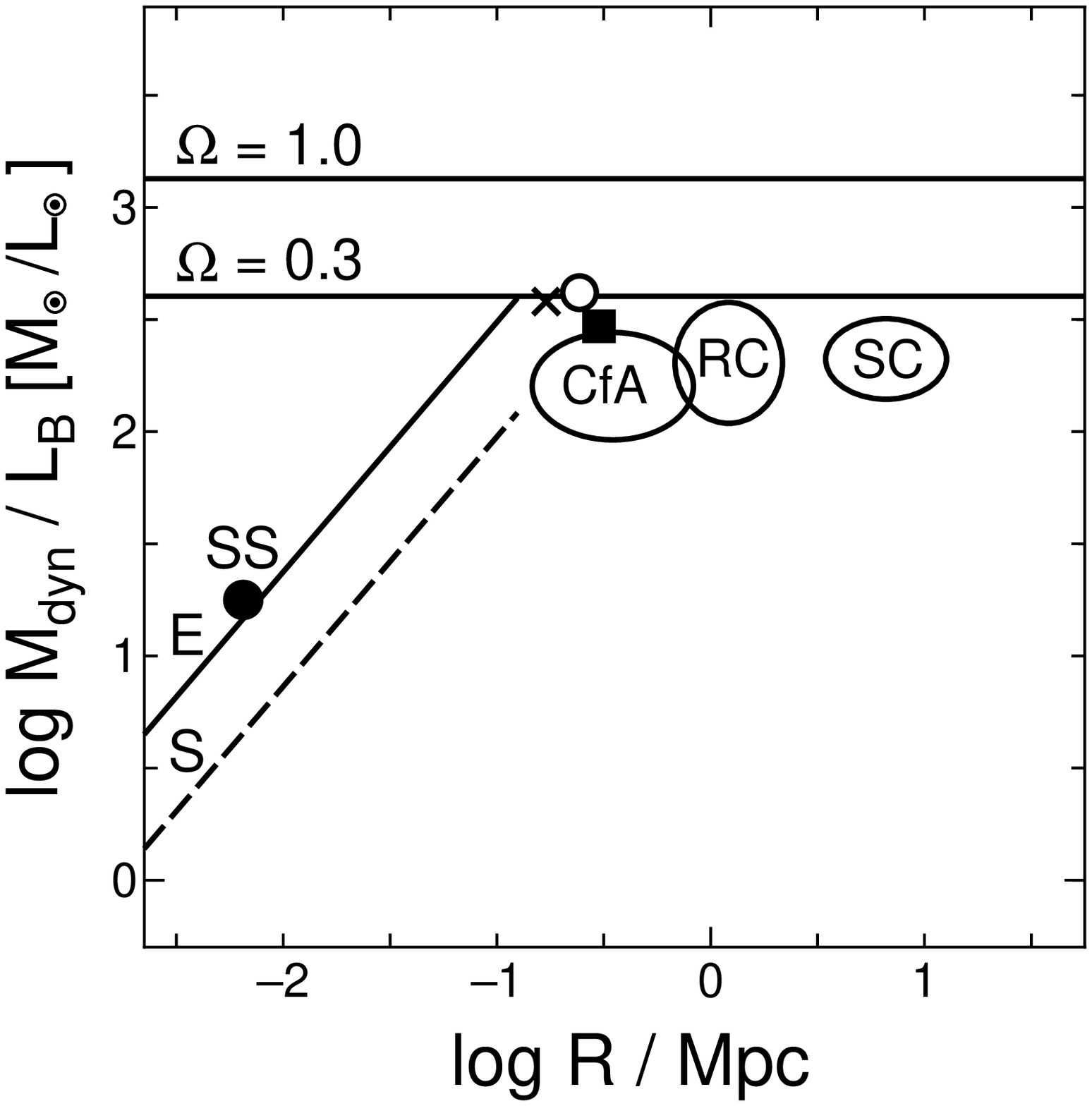}
\figcaption{
The relationship between the mass-to-optical $B$-band luminosity ratio 
and the size from Bahcall et al. (1995). The solid line and dashed line 
indicate the best fit lines for ellipticals and for spirals, 
respectively. The filled circle indicates SS. The open circle means 
NGC 1132. The filled square means NGC 4636. The cross indicates 
RXJ 1340.6$+$4018.
\label{fig8}
}
\end{center}

Finally, we investigate the relationship between the mass-to-light
ratio ($M_{\rm tot}/L_B$ where $L_B$ is the absolute $B$ luminosity)
and  the radius ($R$) of SS and compare it with those of the  
candidate fossil groups and typical elliptical galaxies studied
by Bahcall et al. (1995). 
The basic properties of the ellipticals with large-scale X-ray halos  
are listed in Table 5 taken from Ponman et al. (1994), Matsushita 
et al. (1998), and Mulchaey \& Zabludoff (1999). 
Summing the blue luminosities of the four member galaxies, we obtain
a total blue luminosity for SS,
$L_{B, {\rm tot}} \simeq 2.0 \times 10^{10}h^{-2} L_{\odot}$.
We thus obtain a mass-to-light ratio for SS,
$M/L_B  \simeq 18 h M_{\odot}/L_{\odot}$. 
The linear size of SS is $R \simeq 8 h^{-1}$ kpc
which is the radius of a circle including the center of the four 
member galaxies. We also show these quantities in Table 5. 

It seems worthwhile noting that the relationship between $M/L_B$ and $R$
for SS follows well that for elliptical galaxies (Bahcall et al. 1995: 
see Figure 4). We therefore suspect that the eventual merger remnant of
SS may likely evolve into an ordinary elliptical galaxy. 

\vspace{1ex}

We would like to thank an anonymous referee for his/her useful
comments and suggestions.
We would like to thank the staff members of the
Okayama Astrophysical Observatory and 
the UH 2.2 m telescope for their kind assistance during 
our observations. We thank Harald Ebeling and Makoto Hattori for 
useful discussion on the X-ray data discussed in this paper.
We also thank Richard Wainscoat and Shinki Oyabu for their kind help on
photometric calibration and Youichi Ohyama for his kind help
during the course of this study.
T. M. is supported by JSPS. 
This work was supported in part by the Ministry of Education, Science,
Sports and Culture in Japan under Grant Nos. 07055044, 10044052, 
and 10304013.


\end{document}